\documentclass[conference]{IEEEtran}
\IEEEoverridecommandlockouts
\usepackage{cite}
\usepackage{amsmath,amssymb,amsfonts}
\usepackage{algorithmic}
\usepackage{graphicx}
\usepackage{textcomp}
\usepackage{stfloats,mathcomSTEv4}

\usepackage{cite}

\usepackage{siunitx}
\usepackage{mathtools}
\usepackage{amsmath}
\usepackage{amssymb}
\usepackage{amsmath}

\usepackage{cite}
\usepackage{siunitx}
\usepackage{subfigure}
\usepackage{amsthm}
\usepackage{graphicx}
\usepackage{graphicx}
\usepackage{algorithmic}
\usepackage{graphicx}
\usepackage{algorithm}
\usepackage[utf8]{inputenc}

\usepackage{amsthm}

\usepackage{epstopdf}
\usepackage{mdwmath}
\usepackage{blindtext}
\usepackage{eqparbox}
\usepackage{fixltx2e}

\usepackage{graphicx}
\usepackage{graphicx}

\usepackage{mdwmath}

\usepackage{eqparbox}
\usepackage{xcolor}
\def\BibTeX{{\rm B\kern-.05em{\sc i\kern-.025em b}\kern-.08em
    T\kern-.1667em\lower.7ex\hbox{E}\kern-.125emX}}

    \makeatletter
\newcommand\fs@betterruled{%
  \def\@fs@cfont{\bfseries}\let\@fs@capt\floatc@ruled
  \def\@fs@pre{\vspace*{5pt}\hrule height.8pt depth0pt \kern2pt}%
  \def\@fs@post{\kern2pt\hrule\relax}%
  \def\@fs@mid{\kern2pt\hrule\kern2pt}%
  \let\@fs@iftopcapt\iftrue}
\floatstyle{betterruled}
\restylefloat{algorithm}
\makeatother

\begin{document}

\title{Fairness-Oriented User Association in HetNets Using Bargaining Game Theory}
%
\author{\IEEEauthorblockN{}
\IEEEauthorblockA{\textit{} \\
\textit{}\\
}
\and
\IEEEauthorblockN{Ehsan Sadeghi$^\dagger$, Hamid Behroozi$^\dagger$,		Stefano Rini$^\ddagger$ }
\IEEEauthorblockA{\textit{$^\dagger$ Department of Electrical Engineering, Sharif University of Technology, Tehran, Iran. } \\
\textit{$^\ddagger$ National Chiao Tung University, Hsinchu, Taiwan. }\\
es.ehsan.sadeghi@gmail.com, behroozi@sharif.edu, stefano@nctu.ed.tw
}
\and
\IEEEauthorblockN{}
\IEEEauthorblockA{\textit{} \\
\textit{}\\
}
}
%
\maketitle
\begin{abstract}
%
In this paper, the user association and resource allocation problem  is investigated for a two-tier HetNet consisting of one macro Base Station (BS) and a number of pico BSs.
The effectiveness of user association to BSs  is evaluated in terms of fairness and load distribution. 
%
In particular, the problem of determining a fair user association is formulated as a 
bargaining game  so that for the Nash Bargaining Solution (NBS) abiding the fairness axioms provides an optimal and fair user association.
%
%
The NBS also yields in a Pareto optimal solution and leads to a proportional fair solution in the proposed HetNet model. 
Additionally, we introduce a novel algorithmic solution in which a new Coalition Generation Algorithm (CGA), called SINR-based CGA, is considered in order to simplify the coalition generation phase.
Our simulation results show the efficiency of the proposed user association scheme in terms of fairness and load distribution among BSs and users. 
%
In particular, we compare the performance of the proposed solution with that of the throughput-oriented scheme in terms of the max-sum-rate scheme and show that the proposed solution yields comparable average data rates and overall sum rate. 
\end{abstract}
%
%
\section{Introduction}
%
The comprehensive deployment of small cells, or heterogeneous network (HetNet), is perhaps one the most promising solutions to address the ever growing demand for higher data rates in wireless networks. 
%
HetNets are based on the coexistence of a Macro Base Station (MBS) and small cells (SCs),  differing in transmission power, coverage area, and costs.
The role of the SCs is to bring the network closer to the users, while the MBS's are used to extend coverage and support the SCs' operation. 
%
%
Given the intrinsic difficulty in managing this  two-tier architecture,  user association and resource allocation problems in HetNets  have been widely discussed in the recent literature. 
%
Of particular concern is defining a set of utility metrics that well characterize the actual operation cost of a HetNet, from construction costs, backhaul maintenance,  transmission power, and network resources.  
%
%
%
Some authors focus  purely  on throughput-oriented schemes \cite{myref3, myref4}
while other study fairness-oriented schemes \cite{myref5, myref7, myref8, myref10}.
%
%
Throughput-oriented schemes present a series of disadvantages which are well-understood in the literature: from offering unbalanced data rates and BS loads, to neglecting infrastructural constraints 
To overcome the limitation of throughput-oriented schemes,  various game-theoretic approaches have been considered in the literature 
to address the user association problem \cite{myref5, myref6, myref7, myref8, myref10}. 
%
Among  the game-theoretical tools utilized by these authors, bargaining game theory and its Nash Bargaining Solution (NBS) have emerged as a powerful cooperative game-theoretical tools for efficiently allocating the network resources. 
%
%
%
%
This approach was initially investigated in \cite{myref7}, where it is shown how NBS leads to a proportional fair association. 
In \cite{myref8}, the NBS is utilized for user association and resource allocation in multi-user MIMO OFDMA broadcast system over single-cell area.
%
%
Generally speaking, determining the NBS is computationally complex: hence, algorithmic solutions and approaches are designed to reduce NBS's achievement complexity.
The two-band partition method is first proposed by \cite{myref11} as a low complexity algorithm to divide the number of frequency bins between two users, and its concept is widely utilized in many applications. 
For instance, in \cite{myref10}, the two-band partition method is adopted, and a CGA based on the Hungarian method is provided to achieve the NBS in a single cell OFDMA system channel allocation. 
The same approach as \cite{myref10} is also employed by the authors of \cite{myref6} to solve the NBS in HetNets equipped with simple transceivers in both MBS and Pico Base Stations (PBSs) with different transmitted power. 

\subsubsection*{Contributions}
In this paper, we adopt the NBS as a key tool to address the user association problem in HetNets. 
We believe that Pareto optimality and proportional fairness (PF) of NBS are key tools to achieve fairness and improve load distribution in HetNets. 
%
%
%
Inspired by \cite{myref6}, we choose BSs as competing players in NBS to provide fair utility distribution among BSs: this approach helps event load distribution  and promotes fairness. 
%
The minimum data rate for each user is also set to ensure minimum service requirements and Quality of Service (QoS) for users.
Our main contribution is to analyze fairness in HetNets by solving the NBS optimization problem: we reduce the NBS problem into a convex optimization problem and propose an algorithmic solution. 
Successively, we modify the two-band partition method \cite{myref11}, and rewrite it for our convexified problem in a two-user case. 
Finally, we design a new Coalition Generation Algorithm (CGA) for the multi-user case, the SINR-based CGA, and solve the user association problem using two-band partition and SINR-based CGA. 
We term the overall  scheme as  ``SINR-based Coalition Generation Algorithm (SCGA) NBS scheme (SCGA-NBS)''. 
%
Our major contributions to the work presented in \cite{myref6} are formulating a convex problem, considering NBS constraints, and designing an efficient algorithm to analyze multiple BSs scenario.
Additionally, we compare our SCGA-NBS scheme and throughput-oriented approach in terms of fairness, data rate, load distribution, and convergence time.
We show that our proposed SCGA-NBS scheme outperforms previous schemes as well as throughput-oriented scheme. 
In particular, we prove that the convergence times of our scheme is strictly better than that of the throughput-oriented approach ( a.k.a. max-sum-rate scheme). 
%

\section{System Model}
We consider two-tier HetNet comprised of one MBS and $B-1$ PBSs.
The MBS is denoted as $BS_{1}$, while the PBS as  $BS_{b}$ with $b \in \{ 2, \ldots , B\}$
%
%
The BSs aim to serve $N$ users denoted by $U_{n}$ where $n \in \{1, \ldots N\} \triangleq [N]$.
%
%
The user association is the downlink scenario is expressed through the association index $x_{bn} \in \{0,1\}$ for $b \in [B]$, $n \in [N]$: if $x_{bn}=1$ indicates that the 
the $n^{\rm th}$ user is connected to the $b^{\rm th}$ BS.
%
%
%
The user association matrix $\Xv\in \{0,1\}^{B \times N}$ is define as $\Xv_{bn}=x_{bn}$.
%
%
Every user is served by a single BS, so that each column of $\Xv$ contains a single one.
The effective load, which represents how many users each BS serves, is defined by $L_{b}=\sum_{n \in [N]}x_{bn}$.
Transmissions between BS and the users employ Frequency Division Multiple Access (FDMA): accordingly
the SINR for the $n^{\rm th}$ user located in the $b^{\rm th}$ BS is obtained as:
\begin{align}\label{eq.2}
\text{SINR}_{bn} = \dfrac{ P_{b}\;G_{bn}}{   \underset{i \in [B],\,i \ne b}{\sum} \;P_{i}\;G_{in}+ \sigma^{2} \left( \frac{W}{L_{b}} \right)    } ,
\end{align}
where $P_{b}$\footnote{The same amount of power is assigned to each user to analyze and compare costs and utility of each BS.} is the transmitted power of $BS_{b}$,  $G_{bn}$ represents channel power gain between $BS_{b}$ and $U_{n}$,
$W$ corresponds to the bandwidth which every BS utilizes, and $\sigma^{2}$ indicates the variance of the additive white Gaussian noise. 
We assume that $\sigma^{2}$ is expressed in [dbm/Hz] as it is multiplied by each user's occupied bandwidth,  $ W/L_{b}$: 
accordingly the achievable rate of $n^{\rm th}$ user connected to $b^{\rm th}$ BS can be written as:
\begin{align}\label{eq.3}
r_{bn}=\dfrac{W}{L_{b}}\;\log_{2}(1+\text{SINR}_{bn}).
\end{align}

Let us we next define a fair user association among BSs: a fair user association among BSs not only leads in a simpler backhaul infrastructure for BSs but also can ensure all users to be served fairly.
In this sense, we define the total achievable utility of $BS_{b}$ as:
\begin{align}\label{eq.5}
U_{b}=\sum_{n \in [N]}x_{bn}\,u_{bn}, \quad  u_{bn}=\;\ln \lb \dfrac{r_{bn}}{r_{bn}^{\min}} \rb,
\end{align}
where  $u_{bn}$ is defined as the  $b^{\rm th}$ BS utility depending on the $n^{\rm th}$ user's  achievable rate $r_{bn}$.
Note that $u_{bn}$ can be interpreted as a service gained by $n^{\rm th}$ user connected to $b^{\rm th}$ BS, and this service brings benefits to the serving BS. Our goal is to establish a fair user association based on game theory, where we consider each BS as one of the game players competing to increase its own utility.

\section{Nash Bargaining Solution}
Let us next more formally connect the fairness in the user association problem to the NBS.

\subsection{Proposed Nash Bargaining Solution Based Model}\label{PNBS}
In our model, we consider BSs as  game players in the bargaining game problem.
They compete in increasing their utility as provided in \eqref{eq.5}. In general,  a bargaining game is defined by $( \mathcal{S}, \bar{d})$, where $U=(U_{1}, U_{2}, \ldots, U_{n})$ is a utility vector of the competing players, $\mathcal{S}$ is the feasible set of outcomes, and $\bar{d}=(U_{1}^{*}, U_{2}^{*}, \ldots, U_{n}^{*})$ represents disagreement point, or the minimum achievable profits of the bargaining game.
In a game theoretical setting, the NBS is proposed as a solution to a bargaining problem. 
The NBS considers the \emph{fairness axioms}, and provides a fair solution, that is a solution which is Pareto optimal. 
Also, the NBS can sometimes lead to Proportional Fair (PF) point. 
In the following, we rewrite the productive problem, which should be solved to find NBS for our problem, by considering a relaxation in which the user association matrix $\Xv$ has positive entries smaller than one, that is:
%
\begin{align}
\begin{matrix}
\underset{\Xv}{\max} & U_{\rm Pr}=\prod_{b \in [B]}\;(U_{b}-U_{b}^{*}), \\ \label{eq.7}
&\\
$s.t.$ & U_{b}\geq U_{b}^{*},\;\forall  \ b \in [B]\\ 
& \quad \quad \quad \ \ \ \ \ \   x_{bn}\in [0,1],\;\forall \ b \in [B], n \in [N].
\end{matrix}
\end{align}
Note that the feasible set $\mathcal{S}$ for the optimization problem in \eqref{eq.7}, $(U_{1}, U_{2}, \ldots, U_{N})$, forms a continuous, convex, and compact subset of $\Rbb^{N}$.
This is due to the relaxation and the logarithmic utility function which are the necessary condition for the feasible set for applying NBS.
Substituting \eqref{eq.5} in \eqref{eq.7} leads to a nonconvex problem: for this reason we  modify the objective of the optimization problem in \eqref{eq.7} as:
\begin{align}\label{eq.8}
\Ut_{\rm Pr}=\ln (U_{\rm Pr}) =\sum_{b \in [B]}\ln \left(U_{b}-U_{b}^{*}\right).
\end{align}

This provides us an equivalent optimization problem which leads to the NBS as well:
\begin{align}\label{eq.9}
\begin{matrix}
\underset{\Xv}{\max} & \sum_{b \in [B]} \ln \left(U_{b}\right), \\
&\\ 
$s.t$. & \hspace{-4em} U_{b} \geq 0,\; \forall  \ b \in [B]\\ 
& \quad \quad \quad  x_{bn} \in [0,1],\; \forall  \ b \in [B],\;\forall \ n  \in [N]\\
& \sum_{b \in [B]}x_{bn}=1,\; \forall \ n  \in [N].
\end{matrix}
\end{align}
%
Note that the minimum required utility is set to be zero, i.e. $\bar{d}=(0, \ldots, 0)$. By setting $\bar{d}=(0, \ldots, 0)$, or $U_{b}^{*}=\underline{0},\;\forall  \ b$, it can be shown that this maximization leads to the PF solution which results in a PF allocation among BSs' utilities. 
Therefore, the NBS gives both a fair solution based on its axioms as well as the PF solution. 
Consider $r_{bn}$ in \eqref{eq.3}: we see that $r_{bn}$ is a concave function of the  noise variance and utilized bandwidth.
%
Composing this concave function with concave and none-decreasing logarithm function leads to a concave $u_{bn}$ and $U_{b}$. 
Consequently, using the above modifications of the original optimization problem results in the equivalent problem of \eqref{eq.9}, which is concave and easier to solve. 
In the next session, we introduce an algorithm to determine the solution to the optimization problem  in \eqref{eq.9}.

\subsection{SINR-based Coalition Generation Algorithm (SCGA) NBS scheme}
%
%
Coalitional structures can be applied to the problem in \eqref{eq.9} to determine the NBS efficiently.
Here, we adopt the two-band partition method to solve the bargaining problem in two-player coalitions and design SINR-based CGA to generate two-player coalitions in a simple and fast way.
Algorithmic approaches to find the NBS depend mainly on the initialization phase and the coalition generation. 
%
%
In our algorithm, we present a rather interesting approach to ensure an effective initialization. 
In the \textbf{initialization phase}, we associate users with the BS enjoying better SINR: this procedure leads to a greater Nash product before bargaining; hence, it increases our chance to reach the NBS.
Usually, MBS enjoys higher transmit power. This may lead to a better SINR between users and MBS; consequently, an unfair initialization may take place in two cases. 
First, MBS may be initialized with a large number of users, while PBSs are not.
Second, MBS may be deprived of the appropriate number of users in comparison with PBSs, even though MBS can result in a better data rate, utility, and, subsequently, a bigger Nash production before bargaining.
To reach a fair initialization, we define ``initialized effective load'' for each BS representing the maximum possible number of users associated with each BS in the initialization phase. We consider $\alpha \frac{N}{B}$ and $\frac{B-\alpha}{B-1}\frac{N}{B}$ as ``initialized effective loads'' for MBS and PBs respectively, where $\alpha$ is defined as a proportion to search for all possible ``initialized effective load'' in order to reach a bigger Nash product before bargaining. 
Note that we consider $\alpha \geq 1$, since MBS is supposed to serve more users to reach a greater Nash production before bargaining, and no BS experiences zero effective load.
 The initialization phase is presented in {Alg.\ref{alg3}}.

\noindent
$\bullet\;$ \textbf{Two-player Case:}  We adopt the two-band partition scheme, which is proposed in \cite{myref11} as a key to allocating frequency bins to users in a low complex way. First of all, we will investigate the NBS problem in the two-player case using the two-band partition method, i.e., $\mathcal{B}=\left\{0, 1\right\}$, where the NBS problem will be more simplified.
By virtue of the convex and equivalent problem in \eqref{eq.9}, the dual Lagrangian problem is available, and the duality gap is zero, which is neglected in some researches. With this in mind, by applying Karush–Kuhn–Tucker (KKT) condition, derivations of Lagrangian can be written, i.e. $\dfrac{\partial \mathcal{L}}{\partial x_{b'n'}},\; \forall  \ b' \in \mathcal{B},\;\forall \ n ' \in \mathcal{N}$, where $\mathcal{L}$ stands for Lagrangian.
Assuming the positivity of BSs' utilities and eliminating relaxation constraints simplify our problem.
The problem is further simplified in the two-player case.
By equating Lagrangian derivations of $n^{\rm th}$ user, the following Equations are derived for every single user:
\ean{
%
\dfrac{\sum_{n'\in N} x_{1n'}r_{1n'}^{\min}}{ U_1 L_{1}^{2}} -\f {u_{1n}}{U_1}  
= \dfrac{\sum_{n'\in N}x_{2n'}r_{2n'}^{\min}}{U_2 L_{2}^{2}} -\f {u_{2n}}{U_2},
}
for all $n  \in [N]$.
Usually, a minimum data rate is considered to be constant for all users, i.e., $r_{1n'}^{\min}=r_{2n'}^{\min}=r^{\min}$. 
As a result, by setting $\Qcal_{1}=1/U_{1}$ and $\Qcal_{2}=1 / U_{2} $, we thus  obtain:
\begin{align}\label{eq.13}
\dfrac{{\Qcal}_{1}\;r^{\min}}{L_{1}} -\;{\Qcal}_{1}\,\,u_{1n}\;=\dfrac{{\Qcal}_{2}\; r^{\min}}{L_{2}} -\;{\Qcal}_{2}\,\, u_{2n}.
\end{align}
\begin{algorithm}[t]
\caption{SCGA-NBS scheme- Initialization Phase}\label{alg3}
\begin{algorithmic}[1]
\renewcommand{\algorithmicrequire}{\textbf{Input:}}
\REQUIRE $\alpha$ 
\STATE Set $\alpha\frac{N}{B}$ as an initialized effective load$^{(*)}$ of MBS and set $\frac{B-\alpha}{B-1}\frac{N}{B}$ as an initialized effective load$^{(*)}$ of PBSs
\renewcommand{\algorithmicrequire}{Initial association of the users based on their SINR } 
\FOR {$m=1$ to $N$}
\STATE  Form matrix \textbf{SINR}=[${\rm sinr}_{bn}$]$_{B \times N}$, where ${\rm sinr}_{bn}$ shows SINR between b$^{th}$ BS and n$^{th}$ user
\STATE Find the biggest element of matrix \textbf{SINR} and save its BS and user numbers, i.e. column and row indices, (b$'$, n$'$), respectively
\STATE Associate $n'^{\rm th}$ user with b$^{'\,th}$ BS initially and equal $n'^{\rm th}$ column of \textbf{SINR} matrix to a large negative number
\STATE Check each BS's initialized effective load constraint, and if it is not satisfied, equal the row of that BS to a large negative number in the matrix \textbf{SINR}
\ENDFOR
\STATE Save initial associations in the \textbf{X}$_{0}$ matrix as an initialization matrix
\renewcommand{\algorithmicensure}{\textbf{Output:}}
\ENSURE \textbf{X}$_{0}$ (Initialization matrix)\\
{ \small $^{(*)}$ Assumed to be an integer number.}
\end{algorithmic} 
\vspace{-0.15cm}
\end{algorithm}
The left-hand side of \eqref{eq.13} can be interpreted as the first BS's marginal utility for n$^{th}$ user. Similarly, the right-hand side of \eqref{eq.13} can be interpreted as the second BS's marginal utility for n$^{th}$ user.
Based on two-band partition method presented in \cite{myref11}, we define $\Fcal$ function, $\Fcal(y)=y+\alpha$, which represents marginal and partial benefits of user association with either $BS_{1}$ or $BS_{2}$ for each user.
In $\Fcal$, $y=\Qcal_{1}u_{1n}-\Qcal_{2}u_{2n}$ and $\alpha= (\Qcal_{2}r^{\min}/L_{2}) -(\Qcal_{1}r^{\min}/L_{1})$. 
In proposed algorithm, an iterative approach is considered; therefore, we suppose the association procedure, or initialization phase is done, and $\alpha$, $\Qcal_{1}$, $\Qcal_{2}$, $L_{1}$, as well as $L_{2}$ are constant.
Hence, variables in $\Fcal$ function are $u_{1n}$ and $u_{2n}$. 
$\Fcal$ function values represent how much benefit each BS gets from $n^{\rm th}$ user association.
With this interpretation in mind and the fact that $\Fcal$ is formed by the difference of each BS's marginal benefit, the sign of $\Fcal$ is a brilliant indicator in two-band partition method.
All in all, when $\Fcal(y) > 0$, it is more beneficial for $n^{\rm th}$ user to be associated with $BS_{1}$, and when $\Fcal(y) < 0$, it is more beneficial for $n^{\rm th}$ user to be associated with $BS_{2}$. Note that $\Fcal(y) = 0$ shows there is no difference between being connected to either 1st BS or 2nd BS. 
Finding $\Fcal$ function values for each user helps us to understand how much willing each user is to be associated with 1st BS or 2nd BS. Therefore, we arrange users' $\Fcal$ function values in a decreasing order and change users' numbers based on these arrangements, where bigger ones are more interested in 1st BS, and lower ones are more interested in 2nd BS. Next steps consisting of choosing the right users for each BS, as well as the whole two-band partition based method, is described in {Alg. \ref{alg1}}. 
\begin{algorithm}
\caption{Two-Band  Partition  NBS  (2BP-NBS)  Algorithm  }\label{alg1}
\begin{algorithmic}[1]
\renewcommand{\algorithmicrequire}{\textbf{I. Initialization:}} 
\REQUIRE 
\STATE \,\,\textit{Search for an initialization value of matrix $\Xv$ which satisfies $L_{b} \neq 0,\,b=\left \{1,2\right \}$, and bargaining game feasibility constraints, i.e., $U_{1}, U_{2} \geq 0$}
\STATE \,\,\textit{Calculate $\Qcal_{b}$ and $L_{b},\,b=\left \{1,2\right \}$, using the initialized \textbf{X} }
\STATE \,\,\textit{Set the counter i=1 and $U_{\max}(0)=0$}
\renewcommand{\algorithmicrequire}{\textbf{II. Sort users:}} 
\REQUIRE 
\STATE Reordering users' index in a decreasing way based on their $\Fcal$ function values
\renewcommand{\algorithmicrequire}{\textbf{III. Choosing the right partition for user association:}} 
\REQUIRE 
\FOR {$m=1$ to $N-1$}
\STATE Associate $1^{st}$ with $m^{th}$ user with $BS_{1}$ 
\STATE Associate $(m+1)^{th}$ with $n^{\rm th}$ user with $BS_{2}$
\STATE Calculate $U(m)=U_{1}^{m}   U_{2}^{m}$, where $U(m)$ is the Nash product with the mentioned association, and $U_{b}^{m}, b=\left \{ 1, 2\right \}$ show utilities of BSs based on \eqref{eq.5}
\STATE Check feasibility constraints, i.e. $U_{1}^{m}, U_{2}^{m} \geq 0$, otherwise set $U(m)$ equal to a large negative number
\ENDFOR
\renewcommand{\algorithmicrequire}{\textbf{IV. Update \textbf{X}:}} 
\REQUIRE 
\STATE Select the biggest existing $U(m)$ as the chosen one by two-band partition method and store in $U_{\max}(i)$
\STATE Fix this association and update the association matrix denoted by \textbf{X} 
\renewcommand{\algorithmicrequire}{\textbf{V. Ending criteria }} 
\REQUIRE 
\IF {$U_{\max}(i)\leq U_{\max}(i-1)$}
\STATE Algorithm is converged, and the final association matrix is the $(i-1)^{th}$ associated \textbf{X}
\STATE \textbf{else if}
\STATE Consider the updated association matrix of $i^{th}$ counter as an initialization point
\STATE $i=i+1$
\STATE Back to part II with the new initialized \textbf{X}
\ENDIF
\end{algorithmic} 
\vspace{-0.1cm}
\end{algorithm}

\noindent
$\bullet\; \textbf{Coalition Generation Phase:}$
In the proposed SCGA-NBS scheme, we design an SINR-based CGA to find the most appropriate two-player coalitions in a faster as well as simpler approach and utilize the two-band partition method to solve the NBS in each coalition.
The initialization phase provides matrix \textbf{X}$_{0}$=[x$^{0}_{bn}$]$_{B\times N}$, to search for the best two-player coalition which converges to a maximum of Nash product, i.e. the NBS, where $\left(BS_{i},BS_{j}\right)$ indicates coalition between i$^{th}$ BS and j$^{th}$ BS. 
In each coalition, e.g. $\left(BS_{i},BS_{j}\right)$, a two-player bargaining game will be done between two BSs over their overall initialized users, found by i$^{th}$ and j$^{th}$ column of the matrix \textbf{X}$_{0}$. 
The coalition generation phase aims at finding the coalition partner which leads to a higher Nash production after bargaining on their overall initialized shared users in each coalition.
 While solving the bargaining problem in each coalition, e.g. $\left(BS_{i},BS_{j}\right)$, initialized shared users in each coalition (which belong to BS$_{i}$ or BS$_{j}$) may be transferred from one BS to another; therefore, SINR between users of BS$_{i}$ and BS$_{i}$'s coalition partner, i.e. BS$_{j}$, is valuable and should be noted in order to form a bigger Nash production. To address this issue we generate a benefit matrix $\Omega$=[$\omega_{ij}$]$_{B\times B}$, where each element of matrix represents the benefit of forming $\left(BS_{i},BS_{j}\right)$ coalition. Each element of matrix $\Omega$ is defined as: 
\begin{align}\label{eq.16}
\omega_{ij}=\sum_{k \in [N]}{\rm SINR}_{ik} \textbf{X}(j,k).
\end{align}

It calculates the sum of channel SINRs between $BS_{i}$ and users of $BS_{j}$ (i.e., users which are initialized with $BS_{j}$ in matrix \textbf{X}$_{0}$ in the initialization phase) with the assumption of being served by $BS_{i}$. Note that $\omega_{ij}=\omega_{ji},\, \forall i,j$, and $\omega_{ij}=0,\, \forall i$. The reason behind this benefit definition is that in the formation of the coalition, we must consider all possible associations as well as SINRs between all users and BSs existing in the same coalition. In other words, SINR values between users and the coalition partner (BS) are important because of possible user transitions during solving the bargaining problem in each coalition among BSs. In order to find the best two-player coalition set a ``coalition assignment index'' in which $c_{ij}$ equals to 1 if $i^{th}$ BS forms coalition with $j^{th}$ BS, otherwise it is equal to 0. $c_{ij}$ forms the coalition assignment matrix $C=[c_{ij}]_{B\times B}$, and $c_{ii}=0,\,\forall i$. Note that matrix C indicates the final best possible two-player coalition which should be found through CGA. Our proposed CGA defines a coalition association matrix \textbf{C}=[c$_{ij}$]$_{B\times B}$ using the matrix $\Omega$. Therefore, matrix \textbf{C} determines which BSs should be the coalition partner based on the SINR-based CGA. In this regard, the benefit matrix definition helps us to find the most appropriate partner for every single BS as it is mentioned in {Alg. \ref{alg4}}, which presents our SINR-based CGA.

\noindent
$\bullet\; \textbf{Proposed SCGA-NBS Scheme Overview:}$
Now, we are ready to introduce our proposed NBS scheme, which is shortly presented in {Alg. \ref{alg5}}. 
Note that we consider different values of ``initialized effective load'' in order to find the biggest Nash production in the initialization phase and consider all possible initialization states. Therefore, $\alpha$ varies between 1 to $\beta = \frac{B(N-B-1)}{N}$where $\alpha=1$ shows the equal initialized number of users for MBS in comparison with PBSs, and $\alpha=\beta$ shows the minimum number of the users initialized with PBSs, i.e., one user for each PBS.
\begin{algorithm}
\caption{SCGA-NBS scheme- Coalition Generation Phase}\label{alg4}
\begin{algorithmic}[1]
\renewcommand{\algorithmicrequire}{\textbf{Input:}}
\REQUIRE \textbf{X$_{0}$} (Initialization matrix) 
\STATE Form the benefit matrix \textbf{$\Omega$} by calculating \eqref{eq.16} using the initialization matrix \textbf{X$_{0}$}
\STATE Sort the benefit matrix elements from largest to smallest 
\STATE Find the coalition association matrix \textbf{C}: 
\\ \textit{\textbf{At first}, find the largest element and select two involved BSs as the coalition participants} 
\\ \textit{\textbf{Second}, eliminate other benefit matrix elements related to these two BSs} 
\\ \textit{\textbf{Third}, repeat this procedure till all coalitions are generated} 
\\ \textit{\textbf{At last}, Form the coalition association matrix \textbf{C}} 
\renewcommand{\algorithmicensure}{\textbf{Output:}}
\ENSURE Coalition association matrix \textbf{C}
\end{algorithmic} 
\end{algorithm}
\begin{algorithm}
\caption{SCGA-NBS scheme overview}\label{alg5}
\begin{algorithmic}[1]
\STATE i=1; $\beta = \frac{B(N-B-1)}{N}$ $^{(*)}$;
\FOR {$\alpha=1:\beta$}
\STATE Run {Alg. \ref{alg3}}
\STATE Run {Alg. \ref{alg4}}
\STATE Bargain users in each two-player coalition using the two-band partition, i.e. {Alg. \ref{alg1}}
\STATE Save Nash product of this stage, i.e. $Nash-product(i)$
\STATE i=i+1;
\ENDFOR
\STATE Find $\max \, ( Nash-product)$ and introduce its coalition association matrix as the final coalition association matrix
\STATE Final coalition association matrix results in the NBS \\
{\small $^{(*)}$ Assumed to be an integer number}
\end{algorithmic} 

\end{algorithm}

{Alg. \ref{alg5}} results in the final coalition association matrix, where the bargaining solution in each coalition is determined; therefore, the final user association and the NBS, i.e., the problem in \eqref{eq.9}, are solved. 
%

\emph{Remark:} \textbf{Throughput-Oriented (Max-Sum-Rate):}
Comparing our fairness-oriented approaches with the throughput-oriented (max-sum-rate) approach, which may result in unfair user association, is beneficial. This throughput-oriented association problem can be formulated as follows:
\begin{align}\label{eq.17}
\begin{matrix}
\underset{\Xv}{\max} & \sum_{i \in [B]} \sum_{j \in [N]} \;x_{ij} r_{ij}, & \\ 
$s.t$ & 0\leq x_{ij}\leq 1,\;\forall  \ {i} ,\forall \ {j},&\\ 
&\sum_{i \in [B]} x_{ij}=1,\;\forall  \ {j}.&
\end{matrix}
\end{align}

Note that our throughput-oriented approach is a max-sum-rate problem. With the same approach taken in NBS-based parts, it is easy to show the above optimization problem is convex; hence, we adopt the interior point optimization algorithm to solve \eqref{eq.17}.
\section{Simulation Results}
\label{sec:Simulation Results}
In this section, we compare the performance of our scheme with the NBS-based scheme in \cite{myref6} and our throughput-oriented scheme.
%
We consider 2-tier HetNet with one MBS at the center of the cell assisted with 4 and 6 PBSs in two different scenarios.
One MBS is considered at the center of the cell area with a $167$m radius, and PBSs symmetrically located along a circle with a $120$m radius are assumed. 
$N$ users are uniformly distributed over the macro cell area.
%
In the simulations, a frequency-flat Rayleigh fading model with path loss is considered, i.e. $h_{i,j} \sim \mathbb{C}\mathcal{N}(0,d^{-\beta}_{i,j})$ where $d_{i,j}$ is the distance between $i^{th}$ BS and $j^{th}$ user, and $\beta$ is the path loss exponent. Other relevant simulation parameters are listed in Table \ref{Table1}.
%
%
\begin{table}[htbp]
\caption{Simulation parameters for Sec. \ref{sec:Simulation Results}}
\vspace{-1.5em}
\begin{center}
\begin{tabular}{|c|c|c|c|}
\hline
\textbf{Parameter}&\textbf{Value}&\textbf{Parameter}&\textbf{Value} \\
\hline 
W & 10 MHz&Noise power & -127 dbm/Hz\\
\hline
PBS transmission power &30 dbm&r$^{\min}$ &100 kbps  \\
\hline
MBS transmission power &46 dbm&$\beta$ &3.5 \\
\hline
\end{tabular}
\vspace{-1.5em}
\label{Table1}
\end{center}
\end{table}
%
%
%
%

\begin{figure*}[t]
\vspace{-0.4em}
\hspace{-1.3em}
\centering
\mbox{
\subfigure[Fairness (Jain Index) among competing players(BSs).\label{fig2}]
{\includegraphics[width=3.5in]{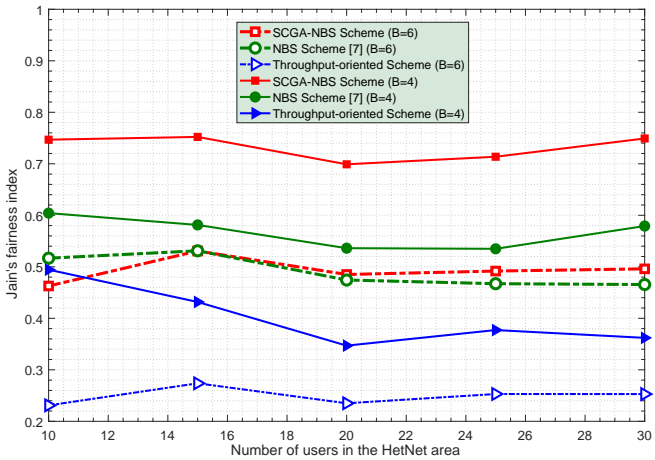}}
\subfigure[Fairness  among users' data rate.\label{fig3}]{\includegraphics[width=3.5in]{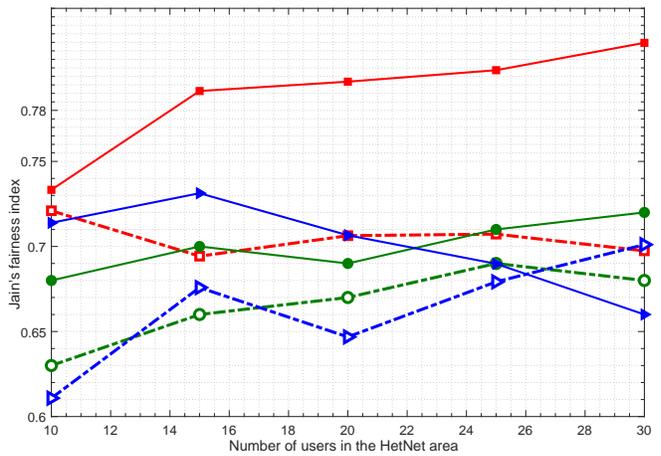}} 
}
\vspace{-.5em}
\caption{Fairness  among competing players (BSs)and users' data rates (Same legend).}
\label{fig23}
\vspace{-1.7em}
\end{figure*}

There are a number of the measures defined in the literature to compare fairness in different association schemes. Jain Index (JI) is one of the quantitative as well as notable measures to evaluate fairness and is defined as follows:
\setlength\abovedisplayskip{1pt}
\begin{align} \label{eq18}
\mathit{f}(x)= \dfrac{\left [ \sum_{i \in [N]} x_{i} \right ]^{2}}{n \sum_{i \in [N]}x_{i}^{2}},
\end{align}
where $0 \leqslant \mathit{f}(x)\leqslant 1 $, $n$ is the total number of shares, and $x_{i}$ presents each person's share. The smaller $\mathit{f}(x)$ is, the more unfair association we have. Note that $\mathit{f}(x)=1$ iff $x_{i}=x,\,\forall i$.
In the proposed game-theoretical model, BSs act as competing players; accordingly, we evaluate JI as the fairness measure for BSs in Fig. \ref{fig2} for 4 and 6 BSs cases simultaneously.

\begin{figure}[t]
\vspace{-.2em}
\includegraphics[width=3.4in]{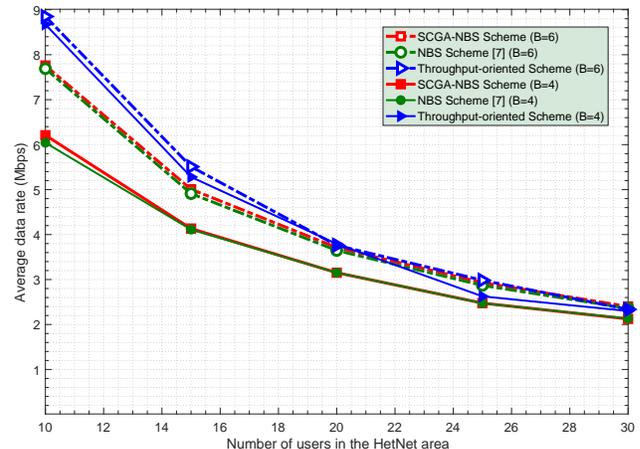}
\vspace{-1em}
\caption {Average data rate (bps) for each user in the HetNets area.}
\vspace{-1.5em}
\label{fig44}
\end{figure}

An unfair user association among BSs may impose a large number of users, or loads on BSs, which violates one of the most critical aims of HetNets, i.e., helping the MBS from being overloaded. Also, the throughput-oriented or unfair association may associate a large number of users with PBSs or the MBS to maximize the sum rate. 
By doing so, the overall data rate may exceed the capacity of the backhaul link of that specific BS (PBS or MBS) while other BSs remain unutilized. 
%
With this in mind, we numerically evaluate the fairness among BSs' utilities and evaluate this fairness of the proposed scheme.
In Fig. \ref{fig2}, it is shown that JI enjoys larger values (more fair solution) for our proposed schemes in comparison with both the scheme presented in \cite{myref6} and the throughput-oriented scheme.\\
\vspace{0em}
Although our problem is formulated to establish fairness among BSs, using numerical results, it is shown that our schemes result in the fair association of data rates among users. In Fig. \ref{fig3}, JI of users' data rates are depicted in the presence of a different number of users. It is shown that our proposed scheme outperforms other schemes and provides a more fair solution to the problem. Considering this, not only is fairness among BSs accomplished, but also fairness among users' data rates is fulfilled. Our NBS-based proposed scheme performs better than the throughput-oriented scheme as well as the NBS scheme in \cite{myref6} and provides more fairness in both BSs' utilities and users' data rates.\\
\vspace{-0em}
Although fairness measures ensure fair association, we should guarantee each user receives its minimum required service (QoS).
In Fig. \ref{fig44}, the average available data rate is found in the presence of a different number of users in 4 and 6 BSs cases. In both cases, the decreasing trend is achieved due to the increasing number of users. Although all NBS-based schemes perform similarly, the proposed SCGA-NBS scheme shows better values.
The throughput-oriented scheme is supposed to maximize the sum rate of the system; hence, it is not surprising to outperform others and also enjoy comparable and close data rates in comparison with the throughput-oriented scheme. $r^{\min}$, which was set in order to satisfy minimum data rate or QoS requirements for each user, is satisfied.\\
%
%
%
%
%
%
%
To investigate backhaul capacity constraints and fairness among BSs, we define a new parameter, termed the Sum Rate Ratio (SRR) as:
\begin{align} \label{eq19}
\text{SRR}=\dfrac{\underset{b\in [B]}{\max}  \  \sum_{n \in [N]}\;x_{bn} r_{bn} }{ \sum_{n \in [N]}\;x_{1n} r_{1n}}.
\end{align}

The numerator in \eqref{eq19} finds a PBS which serves a greater amount of sum rate from associated users, while the denominator represents the sum rate of the users associated with the MBS.
As such, the SRR represents a comparison index for the sum-rate between PBSs and MBS.  
As the MBS enjoys a higher capacity backhaul link in order to send requested data to PBSs, the SRR shows to what extent these infrastructural constraints are met. 
Note that $ { \rm SRR} \in [0,1]$: lower values of the SRRs, from a higher level perspective, indicate the wasting of resources on PBSs while overloading MBS. 
 While the throughput-oriented scheme results in an SRR value less than 0.01, our proposed scheme improves these values to the range $[0.1, 0.2]$.
We observe that the proposed SCGA-NBS scheme presents fair user association between MBS and PBSs as well as among PBSs, while the throughput-oriented scheme fails in both fairness cases.
%
%
%
%
%
%
%
%
%
%
%
%
Indeed the simulation results indicates the proposed NBS-based scheme is successful in achieving the notable overall sum rate of the system by retaining $83-98 \%$ of the overall sum rate achieved by the throughput-oriented scheme. 
Also, our proposed algorithmic solution enjoys lower convergence time in comparison with the throughput-oriented scheme (about five times faster).
\section{Conclusion}
In this paper, we proposed fair user association schemes in HetNets based on  bargaining game theory. Through this formulation, we identify the  user association as the bargaining solution  of a game in which users maximize their utility.
%
Our scheme was presented in order to simplify the coalition generation part by designing the innovative SINR-based approach. This scheme takes advantage of the two-band partition method to achieve the bargaining solution. 
The efficiency of the proposed solution was shown to compare favourably with that of the throughput-oriented approach.
\bibliographystyle{IEEEtran}
\bibliography{IEEEabrvmine}

\end{document}